\documentclass[aps,twocolumn,superscriptaddress]{revtex4-1} 
\usepackage{graphicx}
\usepackage{amsmath}
\usepackage{amssymb}
\usepackage{import}
\usepackage{color}
\usepackage{multirow}
\usepackage{booktabs}
\usepackage{dutchcal}
\usepackage{cleveref}
\crefname{equation}{Eq.}{Eqs.}  
\Crefname{equation}{Equation}{Equations}	
\crefname{figure}{Fig.}{Figs.}
\Crefname{figure}{Figure}{Figures}
\crefname{chapter}{Ch.}{Chs.}
\Crefname{chapter}{Chapter}{Chapters}
\crefname{section}{Sec.}{Secs.}
\Crefname{section}{Section}{Sections}
\crefname{appendix}{App.}{App.}
\Crefname{appendix}{Appendix}{Appendices}	
\crefname{algorithm}{Alg.}{Algs.}
\Crefname{algorithm}{Algorithm}{Algorithm}
\crefname{table}{Tbl.}{Tbls.}
\Crefname{table}{Table}{Tables}

\let\originalleft\left
\let\originalright\right
\renewcommand{\left}{\mathopen{}\mathclose\bgroup\originalleft}
\renewcommand{\right}{\aftergroup\egroup\originalright}
\begin{document}
\title{Finite-difference time-domain simulation of strong-field ionization: Perfectly matched layer approach}
\author{H{\o}gni C. Kamban}
\email{hck@mp.aau.dk}
\affiliation{Department of Materials and Production, Aalborg University, DK-9220 Aalborg \O st, Denmark}
\affiliation{Center for Nanostructured Graphene (CNG), DK-9220 Aalborg \O st, Denmark}
\author{Sigurd S. Christensen}
\affiliation{Department of Materials and Production, Aalborg University, DK-9220 Aalborg \O st, Denmark}
\author{Thomas S\o ndergaard}
\affiliation{Department of Materials and Production, Aalborg University, DK-9220 Aalborg \O st, Denmark}
\author{Thomas G. Pedersen}
\affiliation{Department of Materials and Production, Aalborg University, DK-9220 Aalborg \O st, Denmark}
\affiliation{Center for Nanostructured Graphene (CNG), DK-9220 Aalborg \O st, Denmark}

\date{\today}

\begin{abstract}
A Finite-Difference Time-Domain (FDTD) scheme with Perfectly Matched Layers (PMLs) is considered for solving the time-dependent Schr\"{o}dinger equation, and simulate the ionization of an electron initially bound to a one-dimensional $\delta$-potential, when applying a strong time-oscillating electric field. The performance of PMLs based on different absorption functions are compared, where we find slowly growing functions to be preferable. PMLs are shown to be able to reduce the computational domain, and thus the required numerical resources, by several orders of magnitude. This is demonstrated by testing the proposed method against an FDTD approach without PMLs and a very large computational domain. We further show that PMLs outperform the well known Exterior Complex Scaling (ECS) technique for short-range potentials when implemented in FDTD, though ECS remains superior for long-range potentials. The accuracy of the method is furthermore demonstrated by comparing with known numerical and analytical results for the $\delta$-potential.
\end{abstract}

\maketitle

\section{Introduction}

It has long been established that excitons must be taken into account to accurately describe the optical properties of solids. If the excitons are strongly bound, further complications arise as the excitons must be dissociated into free charge carriers before they can supply an electrical current. Interest in using monolayer transition metal dichalcogenides (TMDs) in electronic devices has increased dramatically during recent years. These materials are known to support strongly bound excitons \cite{Wang2012,Geim2013,Olsen2016}, which dominate their linear and nonlinear optical properties \cite{Wang2012,Ramasubramaniam2012,Qiu2013,Trolle2015}. However, efficient generation of photocurrents in e.g. photodetectors and solar cells require dissociation of excitons into free electrons and holes. It is therefore of significant interest to obtain efficient methods of inducing dissociation of bound excitons in materials such as TMDs. Recently, static in-plane electric fields have proven a promising candidate to aid dissociation \cite{Massicotte2018,Haastrup2016,Pedersen2016ExcitonIonization,Kamban2019}, and the rates induce by such fields are readily calculated by the well known complex scaling method \cite{Balslev1971,Aguilar1971,Haastrup2016,Kamban2019}, or by using a hypergeometric resummation technique \cite{Mera2015}. The situation is not as simple for time-dependent fields, however, which motivates the search for efficient methods of calculating the dissociation rates induced by alternating fields. The problem is remarkably similar to calculating the ionization rate of atoms \cite{Perelomov1966,Keldysh1965,Gersten1974,Tolstikhin2011,Trinh2013,Greenwood1991,Javanainen1988,Plummer1998,Shakeshaft1987,Shakeshaft1988}, for which Floquet theory \cite{Floquet1883} implemented with complex scaling \cite{Holt1983,Maquet1983} has proven very useful. A major drawback with traditional Floquet theory, however, is that it only applies to periodic fields. In the present paper, we seek to develop a different approach based on PMLs. To accurately described TMD excitons, one must use the Keldysh potential \cite{Keldysh1979,Trolle2017}. As this potential is rather complicated, we shall test the method  developed here by calculating the ionization rate of an electron bound to the one-dimensional zero-range potential $-\delta\left(x\right)$.\\
\indent One of the most reliable techniques of obtaining accurate results in intense laser-matter interactions is to propagate the time-dependent Schr\"{o}dinger equation (TDSE) and calculate the relevant observables. Strong electric fields lead to a large probability flux traveling out of the central region occupied by the initially localized wave function, which, of course, is exactly what we mean by ionization. If one simply uses Dirichlet boundary conditions at the edges of the simulation domain, a huge domain is required to avoid reflections of the wave function from the boundary. Several methods have been designed to circumvent this problem, with one of the most common ones being absorbing boundaries outside a specified interior domain. Popular methods include complex absorbing potentials (CAPs) \cite{Riss1996}, absorbing masks \cite{Krause1992}, and ECS \cite{He2007,Tao2009}. The goal of the absorbing layers is to leave the wave function unaltered in the interior region, while absorbing it as it moves out of this region. If it is absorbed sufficiently quickly, then one is able to use Dirichlet boundary conditions at a distance into the layer with minimal flux reaching this point and thus avoiding spurious reflections.

The ECS method has been given a lot of attention in recent years, and rightfully so, as it has been shown to be a very efficient absorber in time-dependent Schr\"{o}dinger problems \cite{McCurdy1991, Scrinzi2010}. A related method that has been given less attention in quantum mechanics is the use of PMLs. The PML method was developed by Berenger for solving Maxwell's equations \cite{Berenger1994}, and has since been used extensively in classical electromagnetism, where PMLs are applied efficiently in FDTD \cite{Inan2011,Taflove2005}, in frequency-domain finite-element \cite{Jin2002}, and in Fourier-series \cite{Zhang2008} approaches. Lu and Zhu additionally proposed a perturbative approach \cite{Lu2005} to deal with undesired effects of the PML when simulating optical wave guides, and PMLs have, furthermore, been utilized to study sound waves \cite{Zuo2017}. Given the success of the PML method in solving problems in electromagnetism, interest in applying it to Schr\"{o}dinger problems has slowly been increasing. Zheng used it to solve the nonlinear Schro\"{o}dinger equation \cite{Zheng2007}, and Nissen and Kreiss have since tried to optimize the PML method for the Schr\"{o}dinger equation with time-independent potentials \cite{Nissen2011}, and have, together with Karlsson, applied it to a reactive scattering problem \cite{Nissen2010}. PMLs have also been applied to time-dependent-density-functional theory (TDDFT)~\cite{Lehtovaara2011}, and the Dirac equation~\cite{Pinaud2015}. It is, however, surprising how comparatively little work has been done on applying PMLs in Schr\"{o}dinger problems, in particular, for explicitly time-dependent problems, such as intense laser-matter interactions.

In the present paper, we develop a method based on a finite-difference time-domain scheme including a PML (FDTD-PML) to describe the ionization of an electron bound by the zero-range $\delta$-potential. This potential has previously been used to, e.g., study the optical response in one-dimensional semiconductors \cite{Pedersen2015} and to model ionization of the $H^-$ ion \cite{Scharf1991}. The convergence of the method will be compared to a standard FDTD scheme using Dirichlet boundary conditions, as well the well known ECS method. We will show that for a short-range potential, the PML method far outperforms ECS when both are implemented as finite-difference schemes. Subsequently, the method will be used to analyze limiting cases, where analytical results can obtained \cite{Francisco1985,Maize2004,Postma1983}. This is done in order to check that the FDTD-PML results remain physical, even though the wave function will be absorbed by the PML. The ionization rate is thereafter calculated as a function of frequency and field strength. Finally, we show that PMLs are not well suited for potentials that reach far into the absorbing layers. 

\section{Electron in a Laser Field}\label{sec:TDSE}

We seek to solve the time-dependent Schr\"{o}dinger equation for an electron, initially bound to a localized potential, perturbed by a monochromatic laser field (atomic units are used throughout)
\begin{align}
	i\frac{d}{dt}\psi\left(\boldsymbol{r},t\right) = \left[-\frac{1}{2}\nabla^2+H_g\left(\boldsymbol{r},t\right)+V\left(\boldsymbol{r}\right)\right]\psi\left(\boldsymbol{r},t\right)\thinspace,
\end{align}
where $H_g$ describes the interaction between the electron and the laser field. Here, the subscript $g$ refers to the gauge, in which the interaction is considered. We shall work only in the dipole approximation such that neither electric fields nor vector potentials have any spatial dependence. This leads to the interaction in the velocity gauge (VG) being
\begin{align}
	H_V = \boldsymbol{p}\cdot\boldsymbol{A}\left(t\right)\thinspace,\label{eq:VGHg}
\end{align}
where $\boldsymbol{p}$ is the momentum operator and $\boldsymbol{A}$ is the vector potential. Note that the usual $A\left(t\right)^2/2$ term has been removed by a unitary transformation. We will consider the monochromatic field defined by 
\begin{align}
\boldsymbol{A}\left(t\right) = A_0\cos\left(\omega t\right)\hat{\boldsymbol{x}}\thinspace.
\end{align}
In the length gauge (LG), the interaction is given in terms of the electric field 
\begin{align}
\mathbcal{E}\left(t\right)=-\frac{\partial \boldsymbol{A}\left(t\right)}{\partial t}=\mathcal{E}_0 \sin\left(\omega t\right)\hat{\boldsymbol{x}}
\end{align}
by
\begin{align}
	H_L = \boldsymbol{r}\cdot \mathbcal{E}\left(t\right)\thinspace.\label{eq:LGHg}
\end{align}
The goal is to be able to reproduce the exact wave function in an interior box $\left|\boldsymbol{r}\right|\leq R_0$ for relevant time periods. That is, we seek to modify the TDSE so that the solution to the modified equation $\psi$ satisfies
\begin{align}
	\psi\left(\boldsymbol{r},t\right) = \psi_{\mathrm{ex}}\left(\boldsymbol{r},t\right) \quad \text{for } \left|\boldsymbol{r}\right|\leq R_0\thinspace,
\end{align}
where $\psi_{\mathrm{ex}}$ is the exact wave function. To be able to quantify the error by a single number, we will use the error measurement introduced by Scrinzi \cite{Scrinzi2010}
\begin{align}
	\sigma\left(R_0\right) = 1- \frac{\left|\left<\psi_{\mathrm{ex}}|\psi\right>_{R_0}\right|^2}{\left<\psi_{\mathrm{ex}}|\psi_{\mathrm{ex}}\right>_{R_0}\left<\psi|\psi\right>_{R_0}}\thinspace,\label{eq:error}
\end{align}
where the scalar product is to be taken in the region $\left|\boldsymbol{r}\right|\leq R_0$, i.e.
\begin{align}
	\left<f|g\right>_{R_0} = \int_{\left|\boldsymbol{r}\right|\leq R_0}f^*\left(\boldsymbol{r}\right)g\left(\boldsymbol{r}\right)d\boldsymbol{r}\thinspace.
\end{align}
\section{Exterior Complex Scaling}\label{sec:ECS}

The literature covering complex scaling is vast (see e.g. \cite{Balslev1971,Aguilar1971,Doolen1974,Ho1981,Reed1982,Bengtsson2008} and references therein), and for this reason we shall only describe briefly the most relevant aspects to the present paper before describing how PMLs are implemented. For simplicity, we will restrict the discussion to one dimension $x$. The simplest form of complex scaling is implemented by scaling the coordinates uniformly according to $x\rightarrow xe^{i\theta}$\thinspace, where $\theta$ will be taken as a purely real number. The motivation is that the outgoing waves $\exp\left(ikx\right)$ become exponentially decaying waves if the rotational angle $\theta$ is chosen large enough. This transformation, referred to as uniform complex scaling (UCS), has been used with great success in finding ionization rates for static electric fields \cite{Herbst1978,Pedersen2016Stark,Massicotte2018} and in solving the TDSE \cite{Bengtsson2008}. In certain situations, however, one may wish to leave the domain untransformed in an interior region and introduce complex scaling only in an outer region. The original motivation was that UCS cannot be used with the Born-Oppenheimer approximation \cite{Simon1979}. Furthermore, when dealing with sufficiently weak electric fields, ionization rates typically become so low that UCS results in a wave function that (numerically) vanishes before it reaches the important region far from the core \cite{Kamban2019}. The ECS \cite{Simon1979,McCurdy2004,McCurdy1991,Rescigno2000} procedure circumvents these problems, and is implemented in one dimension by the transformation
\begin{align}
	x\rightarrow \tilde{x} = \begin{cases}
	x\quad & \mathrm{for }\, \left|x\right|<R_0\\
	e^{i\theta}\left(x\pm R_0\right) \mp R_0 \quad &\mathrm{for }\, \mp x>R_0\thinspace,
	\end{cases}\label{eq:ECStrans}
\end{align}
where $R_0$ is referred to as the scaling radius. This transformation turns outgoing waves into decaying waves in the absorbing layer while leaving them unaffected in the interior region. 

The resulting behaviour of the wave function in the absorbing layer is slightly different in the two gauges. The exponential propagator can be constructed as usual (see Ref. \cite{He2007}), and the perturbing part can be written as
\begin{multline}
	\exp\left(-iH_L\Delta t\right) = \exp\left\{-i\mathcal{E}\left(t\right)\left[\cos\theta\left(x\pm R_0\right)\mp R_0\right]\Delta t\right\}\\
	\times \exp\left\{\mathcal{E}\left(t\right)\sin\theta\left(x\pm R_0\right)\Delta t\right\}\thinspace,\label{eq:expLG}
\end{multline}
in LG and as
\begin{multline}
	\exp\left(-iH_V\Delta t\right) = \exp\left(-i\cos\theta A\left(t\right) p_x \Delta t\right) \\ \times \exp\left(-\sin\theta A\left(t\right)p_x\Delta t\right)\thinspace,\label{eq:expVG}
\end{multline}
in VG. The first terms on the right-hand sides of Eqs. \eqref{eq:expLG} and \eqref{eq:expVG} are oscillatory, while the second terms are either exponentially increasing or exponentially decreasing. In LG, this depends on the sign of the oscillatory field $\mathcal{E}\left(t\right)$ and in VG on the sign of $A\left(t\right)$. Thus, in both cases one may obtain an undesired exponentially increasing behavior in the absorbing layer. Given that the behavior depends on the sign of the field or vector potential, the propagators will oscillate between amplifying and damping the wave function exponentially. In practice, we have found that the exponential behavior outside of the scaling radius is much more apparent in LG than inVG. Furthermore, in LG, the exponential behavior is more volatile for larger $x$, which may lead to numerical instabilities if a wide absorbing layer is desired.
In practice, we have not found these growing terms to cause numerical instabilities for moderate frequencies, while for low frequencies they lead to a numerically diverging wave function. This is in agreement with the observations in Ref. \cite{He2007}. 
\section{Perfectly Matched Layers}\label{sec:PML}

The PML scheme for the TDSE is usually derived by assuming that the potential is both spatially and temporally invariant, and then modal analysis is performed on the Laplace-transformed equation to ensure that the solution decays outside the interior domain, i.e. $\left|x\right|>R_0$ \cite{Zheng2007,Nissen2011,Nissen2010}. The transformation can be formulated as 
\begin{align}
x\to \tilde{x} = \begin{cases}
x\quad & \mathrm{for }\, \left|x\right|< R_0\\
x + i\sigma_0\int^{x}f\left(x'\right)dx' \quad &\mathrm{for }\, \left[x\right]>R_0\thinspace,
\end{cases}\label{eq:PMLxtrans}
\end{align}
where $\sigma_0$ is a constant referred to as the absorption strength and $f$ is the absorption function. The absorption function is zero inside the interior $\left|x\right|\leq R_0$ and positive otherwise. Specific forms will be discussed later. Unlike in ECS, the transformation in \cref{eq:PMLxtrans} is not applied to the potential. Thus, the PML method can be understood as a transformation of the differential operator
\begin{align}
	\frac{\partial}{\partial x} \to c\left(x\right)\frac{\partial}{\partial x}\thinspace,\label{eq:PMLtrans}
\end{align}
where $c\left(x\right) = 1/\left[1 + i\sigma_0f\left(x\right)\right]$\thinspace. The PML equation in one dimension therefore becomes
\begin{align}
	i\frac{\partial }{\partial t}\psi = \left[-\frac{1}{2}c\left(x\right)\frac{\partial}{\partial x}c\left(x\right)\frac{\partial}{\partial x}+H_g\left(x,t\right)+V\left(x\right)\right]\psi\thinspace,\label{eq:PML}
\end{align}
 which coincides with the usual TDSE inside a box of radius $R_0$ as $c\left(\left|x\right|\leq R_0\right)=1$. Note that the momentum operator in $H_V$ (see \cref{eq:VGHg}) is also transformed according to \cref{eq:PMLtrans}. As the transformation is only applied to the spatial derivatives, it is only reasonable to expect Eq. \eqref{eq:PML} to yield a good approximation if $R_0$ is chosen sufficiently large so that the variations in the potential $V\left(x\right)$ in the exterior are negligible. This is the case for any non-zero $R_0$ for the zero-range potential $V\left(x\right)=-\delta\left(x\right)$. However, the interaction in LG effectively modifies the potential so that it includes a linear term which does not vanish outside the interior. One may therefore speculate to which degree \cref{eq:PML} in LG is able to approximate the exact wave function in the interior. Indeed, as we show numerically below, implementing the PML in LG introduces significantly larger errors than in VG. The technical details of implementing both the ECS and PML method can be found in \cref{app:FD}.

\subsection{Absorption function}

It is important that the absorption function $f$ be chosen positive to ensure decay of the wave function as it travels out of the interior domain. Previous choices include low-degree power functions \cite{Zheng2007,Berenger1994} and singular functions \cite{Bermudez2004}. It is interesting to examine whether or not there is a substantial difference between the numerical accuracy obtainable with the different functions. To this end, we will compare four different absorption functions, namely 
\begin{align}
	f\left(y\right) = \Theta\left(y\right)\begin{cases}
		d/\left(d-y+\epsilon\right) -1\\
		y^2\\
		y^3\\
		\tanh\left(2y/d-1\right) -\tanh\left(-1\right)\thinspace,
	\end{cases}
\end{align}
where $y = \left|x\right|-R_0$, $\epsilon$ is some small positive number, and $\Theta$ is a step function equal to unity for $y\geq 0$ and zero otherwise, and $d$ is the width of the absorbing layer. For $\epsilon$, we have used $10^{-4}$, as we have not found the results to be highly dependent on $\epsilon$.

\section{Short-range Potential}\label{sec:SR}
\begin{figure}[t]
	\includegraphics[width=1\columnwidth]{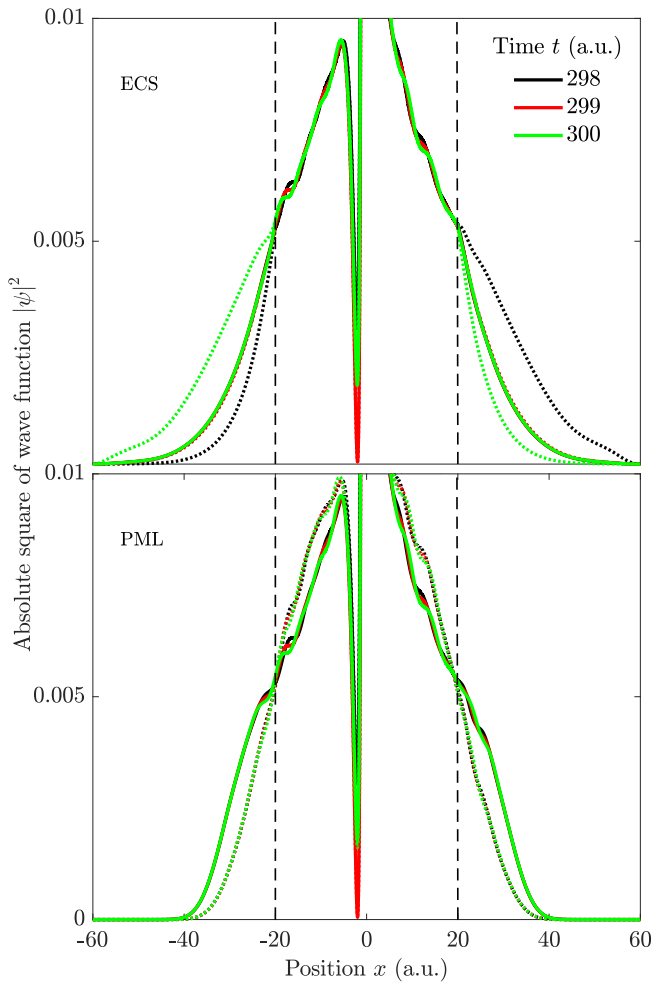}
	\caption{Absolute square of wave functions calculated with ECS (upper) and PML (lower) for three different times with $R_0 = 20$ a.u. (indicated by vertical dashed lines). The solid and dotted lines are calculated in VG and LG, respectively. The field parameters are $\mathcal{E}_0 = 0.1$ a.u. and $\omega=0.52$ a.u., and the field has been turned on smoothly over three optical cycles.}\label{fig:WFSW}
\end{figure}
The first potential we will consider is the short-range potential
\begin{align}
	V\left(x\right)=\begin{cases}
	-\frac{1}{2b} \quad &\mathrm{for}\,\left|x\right|<b\thinspace,\\
	0	\quad &\mathrm{otherwise}\thinspace.\label{eq:srpot}
	\end{cases}
\end{align}
This potential can be seen as a discrete approximation to the zero-range potential $-\delta\left(x\right)$, for which we want to calculate the ionization rate. We have used $b=5\times 10^{-3}$ for the calculations in the present paper, which leads to a ground-state energy of $E_0 = -0.4967\, \mathrm{a.u.}$ (as opposed to $-1/2\, \mathrm{a.u.}$ for the zero-range potential).

\begin{figure}
	\includegraphics[width=1\columnwidth]{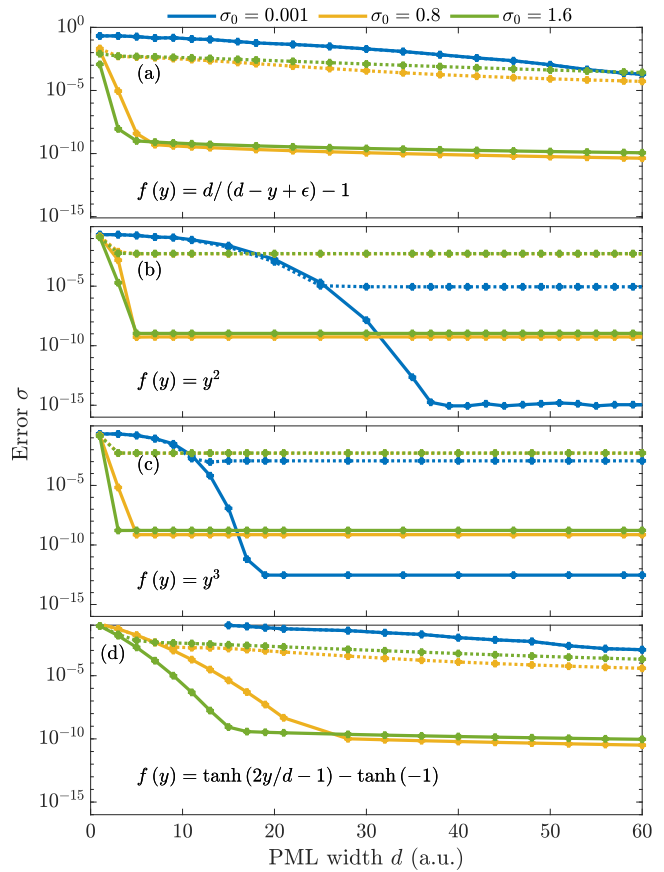}
	\caption{Error at $t = 200$  a.u. as a function of PML width for four different absorption functions and various absorption coefficients. The solid and dotted lines correspond to PML VG and LG, respectively. The scaling radius is set to $R_0 = 20$ a.u.. The field parameters are the same as in \cref{fig:WFSW}. }\label{fig:errorvsPMLSW}
\end{figure}
\Cref{fig:WFSW} shows the absolute square of the wave function calculated with the ECS and the PML method at three different times. The oscillations in LG  outside the scaling radius ($R_0=20\,\mathrm{a.u.}$) that were discussed briefly in the ECS section above are immediately clear. The absolute square of the wave function is, however, graphically indistinguishable in the interior for ECS implemented in LG and VG. For the PML calculations, the results are not as equal-footed. The LG calculation introduces non-negligible reflections leading to a large error inside the interior. This is, of course, what we seek to avoid and thus one must be careful in implementing PML in LG. For VG, the PML and ECS wave functions are indistinguishable for $\left|x\right|\leq R_0$.

\begin{figure}[t]
	\includegraphics[width=1\columnwidth]{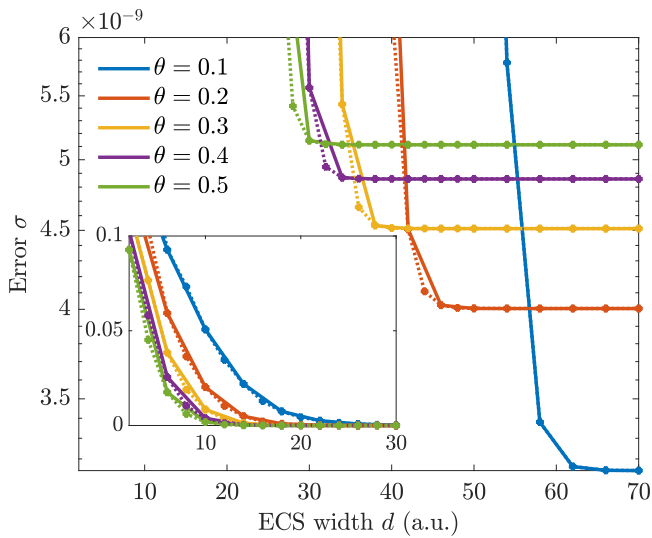}
	\caption{Error at $t=200$  a.u. as a function of ECS width for various angles of rotation. The field parameters are the same as in \cref{fig:WFSW}. The inset shows the behavior for smaller widths. }\label{fig:errorvsECSSW}
\end{figure}
To perform error analysis, we need a reference function. It was obtained by calculating the error in \cref{eq:error} inside $R_0 = 20$ a.u. at $t = 200$ a.u. between two wave functions without any transformation and with Dirichlet boundary conditions set at $x=\pm n\times500$ for $\psi$ and $x=\pm \left(n+1\right)\times500$ a.u. for $\psi_{\mathrm{ex}}$, where $n$ is a positive integer, which was increased by one until the error vanished within numerical precision. This occurred at $n=8$, and thus without any absorbing layer a domain width of at least $8000$ a.u. is needed. To ensure a numerically exact reference function, we have used a domain width of $10000$ a.u. for $\psi_{\mathrm{ex}}$ in the error calculations. In \cref{fig:errorvsPMLSW} we show the error calculated by \cref{eq:error} at $t = 200$ a.u. as a function of the PML width $d$ for the different absorption functions with various absorption strengths $\sigma_0$. It is clear that the PML should be implemented in VG to obtain an accurate wave function in the interior. We also notice that an absorption function that grows slowly leads to a lower error at the cost of slower convergence. The reason is that less of the wave function will be (numerically) reflected upon entering the absorbing layer when the transition is more gradual. The error introduced by the power functions converge to a constant value for a sufficiently large $d$. This indicates that the entire outgoing flux has either been absorbed or reflected from the layer, and further increasing $d$ does not make any difference. It is worth noting that the PML method with a square absorption function (\cref{fig:errorvsPMLSW}(b)) leads to errors of order $10^{-15}$ for an absorbing layer of around $d=40$ a.u.. Thus, the domain width required to obtain an excellent approximation to the exact wave function is $2\left(d+R_0\right)=120$ a.u., significantly lower than the domain width of $8000$ a.u. needed without any absorbing layer. For the nearly singular function and the $\tanh$ function, increasing $d$ leads to an absorption function that grows more slowly. For this reason, the errors they induce do not converge in the same manner as those induced by the power functions. Rather, they continue to decrease as the absorbing layers become wider. 
\begin{figure}[t]
	\includegraphics[width=1\columnwidth]{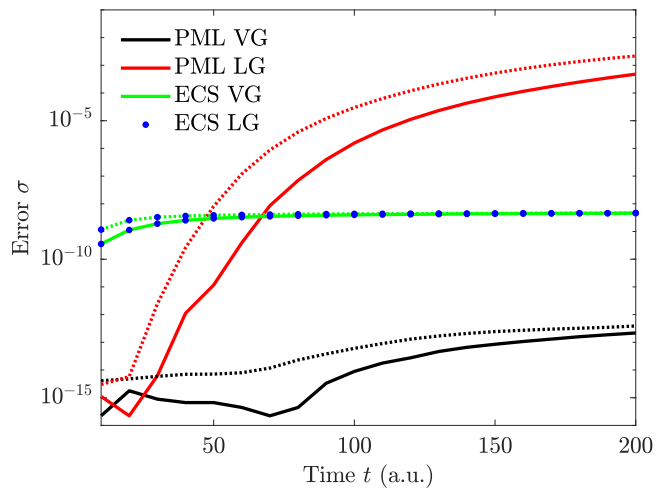}
	\caption{Error as a function of time. The absorbing boundary is located at $R_0=20$ and at $R_0=10$ for the solid and dotted lines, respectively. The PML calculations have been made with $\sigma_0 = 0.001$ and the quadratic absorption function, while the ECS calculations have been made with $\theta = 0.35$. In both cases the absorption width is $d = 40$ a.u.. The field parameters are the same as in \cref{fig:WFSW}. }\label{fig:errorvstimeSW}
\end{figure}
The PML errors should be compared to those introduced by the ECS method, shown in \cref{fig:errorvsECSSW}, for which both LG and VG calculations converge to the same error, and are nearly indistinguishable. A low rotational angle $\theta$ can be seen to introduce lower errors, as less of the wave function will be reflected upon entering the absorbing layer, exactly as with the PML method. By comparing the errors introduced by ECS to those for PMLs with a square absorption function in \cref{fig:errorvsPMLSW}(b), we see that the ECS errors for comparable absorption widths converge to values that are around six orders of magnitude larger. For PMLs with a quadratic absorption function (\cref{fig:errorvsPMLSW}(c)), the ECS errors are around four orders of magnitude larger. This might be due to the poor performance of ECS when implemented in finite-difference schemes \cite{McCurdy2004}.
The same behavior is observed in \cref{fig:errorvstimeSW}, where the error is shown as a function of time. The PML LG calculation introduces much larger errors than the other two methods, and the ECS LG and VG errors are graphically indistinguishable. Again, the PML VG calculation leads to the lowest error by several orders of magnitude. The dotted lines show a scaling radius of $R_0 = 10\,\mathrm{a.u.}$ as opposed to $R_0 = 20\,\mathrm{a.u.}$. As can be seen, reducing the size of the box does not have a large impact on the errors in the interior domain.

%

\subsection{Polarizability and ionization}

\begin{figure}[b]
	\includegraphics[width=1\columnwidth]{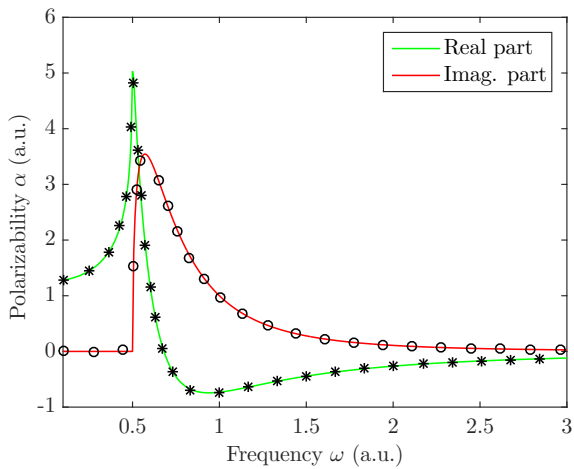}
	\caption{Polarizability calculated from first order perturbation theory (solid lines) and the PML method (markers). A field strength of $\mathcal{E}_0=10^{-6}$ a.u. was used. }\label{fig:pol}
\end{figure}

In the previous section, the error of the wave function inside the box $\left|x\right|<R_0$ was analyzed. If the wave function can be reproduced, then the desirable observables can be calculated, as long as the box size is chosen adequately. While the error measurement defined by \cref{eq:error} is a meaningful parameter, we are unable to directly relate it to physical observables. As an additional check, we therefore demonstrate that we are able to reproduce the frequency dependent polarizability in the weak-field limit. Numerically, the polarizability $\alpha$ can be found by calculating $\left<x\right>/\mathcal{E}_0$ in the weak field limit, and relating it to the real and imaginary part of $\alpha$. Here, $\left<x\right>$ is the average value of $x$. As the field amplitude is extremely low, and $\left<x\right>$ is only needed over a single period, the integral in $\left<x\right>$ can, to an excellent approximation, be restricted to the interior region. An analytical expression can be found for $\alpha$ of the $\delta$-potential ground-state using linear perturbation theory \cite{Postma1983}. It is given by 
\begin{align}
\alpha\left(\omega\right) = \frac{2-\omega^2-\sqrt{1+2\omega}-\sqrt{1-2\omega}}{\omega^4}\thinspace.
\end{align}
As can be seen in \cref{fig:pol}, the PML simulations using a low field strength of $\mathcal{E}_0=10^{-6}$ a.u. are in excellent agreement with the analytical results. 
\begin{figure}
	\includegraphics[width=1\columnwidth]{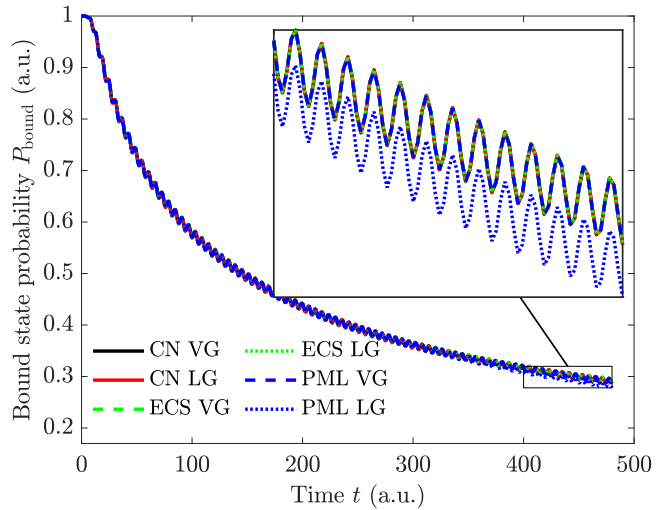}
	\caption{Probability of occupying a bound state as a function of time. The field parameters are the same as in \cref{fig:WFSW} and has been turned on smoothly over one optical cycle. }\label{fig:boundstateprobSW}
\end{figure}

A special interest in the present paper is to obtain the strong-field ionization rate. The probability of occupying a bound state (that is, not being ionized) can be found by
\begin{align}
P_{\mathrm{bound}}\left(t;\mathcal{E}_{0}\right) = \sum_b\left|\left<\varphi_b|\psi_{\mathrm{ex}}\right>\right|^2\thinspace,\label{eq:Pbs}
\end{align}
where the sum is to be taken over all bound states. For numerical calculations, however, one may cut the sum after convergence to a desired number of digits. Let us denote the most delocalized state included in the sum by $\varphi_B$, such that from some number $L$
\begin{align}
	\left|\varphi_B\left(x\right)\right|\geq \left|\varphi_{b}\left(x\right)\right| \quad \text{for } \left|x\right|\geq L\thinspace,
\end{align}
where $b$ refers to all states included. If $L$ is chosen such that $\varphi_B$ is negligible for $\left|x\right|>L$, then all integrals in \cref{eq:Pbs} may be restricted to $\left|x\right|<L$. By choosing the scaling radius $R_0$ to coincide with $L$, we can therefore describe all bound states, as well as obtain an excellent approximation to the wave function, in the interior domain, allowing us to implement \cref{eq:Pbs} in the present approach. For the short-range potential defined by \cref{eq:srpot} there is only one bound state and it decays exponentially for $\left|x\right|>b$. Therefore, a scaling radius of $R_0=20\,\mathrm{a.u.}$ is expected to be adequate. The probability of not being ionized can be seen as a function of time in \cref{fig:boundstateprobSW}, where the ECS and PML calculations are compared to a converged Crank-Nicolson calculation in an untransformed domain. As is evident, the result obtained by the PML method in LG is the only one that can be distinguished from the other ones. This is yet another indication that care must be taken when implementing PMLs in LG.  

\begin{figure}[b]
	\includegraphics[width=1\columnwidth]{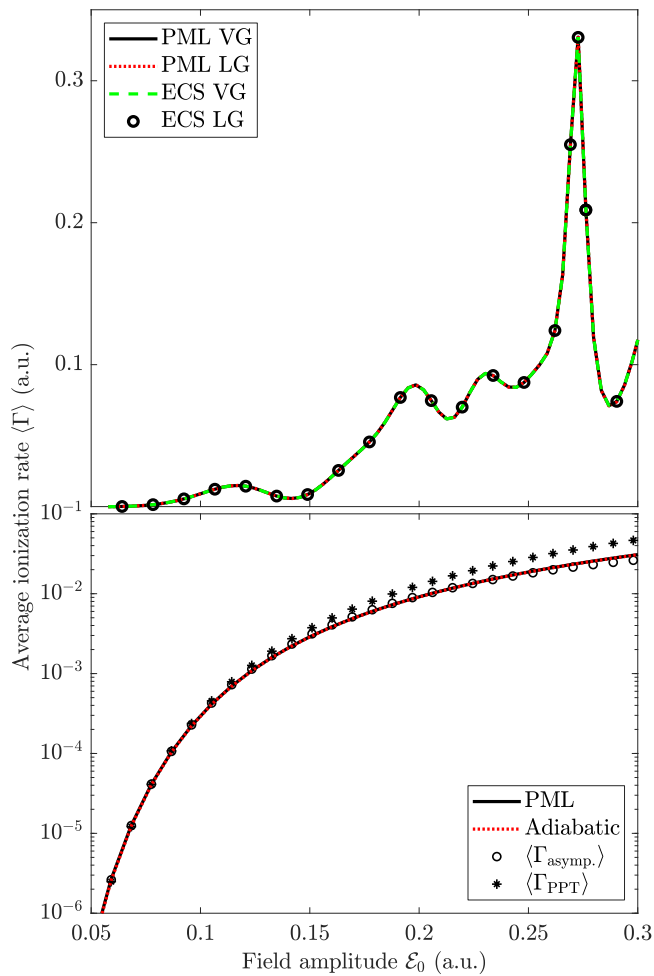}
	\caption{Ionization rate for $\omega = 0.2\,\mathrm{a.u.}$ (upper) and $\omega = 0.01\,\mathrm{a.u.}$ (lower). For the larger frequency, the vector potential has been turned on smoothly over $t=35\,\mathrm{a.u.}$, while for the lower frequency the electric field has been turned on linearly over the same amount of time. The ionization rate in Eq. \eqref{eq:ionizationrate} has been averaged over $\left[4\pi/\omega;8\pi/\omega\right]$ for the larger frequency, and over $\left[\pi/2\omega;5\pi/2\omega\right]$ for the lower frequency.} \label{fig:ionizationrateSW}
\end{figure}

To obtain the time-dependent ionization rate $\Gamma$, we use
\begin{align}
\Gamma\left(t;\mathcal{E}_{0}\right) = -\frac{d}{dt}\ln P_{\mathrm{bound}}\left(t;\mathcal{E}_{0}\right)\thinspace.\label{eq:ionizationrate}
\end{align}
The ionization rate defined by \cref{eq:ionizationrate} will oscillate in time. It is therefore convenient to average the time-dependent ionization rate over a number of periods to remove these oscillations, and thereby obtain a time-independent ionization rate, i.e.
\begin{align}
\left<\Gamma\left(\mathcal{E}_0\right)\right> = \frac{\omega}{2n\pi}\int_{t_0}^{t_0+2n\pi/\omega}\Gamma\left(t;\mathcal{E}_{0}\right) dt\thinspace,\label{eq:average}
\end{align}
where $n$ is the number of periods, and $t_0$ is an initial time taken after the field has been turned on. The ionization rate averaged over two periods can be seen in the upper panel of \cref{fig:ionizationrateSW} for $\omega = 0.2\,\mathrm{a.u.}$. As is evident, the four methods yield identical results. Furthermore, the shape of the ionization rate is consistent with the results for three-photon ionization in Ref. \cite{Scharf1991}.  

In the adiabatic limit, one can obtain analytical results for the ionization rate of the zero-range potential. By setting up the Schr\"{o}dinger equation for a static electric field $\mathcal{E}_{\mathrm{DC}}>0$ and requiring that the wave function becomes an outgoing wave as $x\rightarrow-\infty$ \cite{Francisco1985}, one can obtain the following condition
\begin{align}
\frac{\mathcal{E}_{\mathrm{DC}}^{1/3}}{2^{2/3}\pi}-\text{Ai}\left(\lambda\right)\text{Bi}\left(\lambda\right) = i\text{Ai}^2\left(\lambda\right)\thinspace,\label{eq:Airy}
\end{align}
where $\lambda = -2^{1/3}E\mathcal{E}_{\mathrm{DC}}^{-2/3}$, and $\text{Ai}$ and $\text{Bi}$ are Airy functions of the first and second kind \cite{Abramowitz1972}, respectively. Solving \cref{eq:Airy} numerically one obtains complex energies and the DC ionization rate is then given by $\Gamma_{\mathrm{DC}}\left(\mathcal{E}_{\mathrm{DC}}\right) = -2\text{Im}\left[E\left(\mathcal{E}_{\mathrm{DC}}\right)\right]$ \cite{Francisco1985,Herbst1978}. In the adiabatic regime, the ionization rate by an oscillating monochromatic field is the cycle average of the DC ionization rate corresponding to the instantaneous static electric field at a specific time $\mathcal{E}\left(t\right)$ \cite{Joachain2011}, that is
\begin{align}
\left<\Gamma_{\mathrm{Adiabatic}}\left(\mathcal{E}_0\right)\right> = \frac{\omega}{2\pi}\int_0^{2\pi/\omega}\Gamma_{\mathrm{DC}}\left[\mathcal{E}\left(t\right)\right]dt\thinspace.\label{eq:adint}
\end{align}
As can be seen in the lower panel of \cref{fig:ionizationrateSW}, this adiabatic ionization rate corresponds exceptionally well with the average ionization rate obtained from \cref{eq:average} with $t_0 = \pi/2\omega$ and $n=1$, i.e., $t\in\left[\pi/2\omega;5\pi/2\omega\right]$. It should be noted here that, in this low frequency limit, ECS in both LG and VG, as well as PML in VG, diverge numerically. This phenomenon was briefly discussed above and in more detail in Ref. \cite{He2007}. Thus, the low frequency ionization rate has been calculated implementing PML in LG. This can be done without obtaining significant errors for two reasons: (i) we are only interested in the interval $t<5\pi/2\omega$, for which the PML LG method gives a fairly good approximation to the real wave function for low frequencies, and (ii) the ionization rate is not as sensitive to the errors in the wave functions as the error measurement in \cref{eq:error}.

Two further comparisons are made in the lower panel of \cref{fig:ionizationrateSW}. The first is an analytical approximation to the low frequency ionization rate. This can be obtained by analyzing the asymptotic behavior of the Airy functions. For low field strengths, $\left|\lambda\right|$ tends to infinity. In fact, both the real and imaginary part of $\lambda$ tend to $+\infty$ for $\mathcal{E}_0\to0$. Substituting the asymptotic expressions of the Airy functions \cite{Abramowitz1972} into  \cref{eq:Airy}, one obtains a polynomial in the electric field multiplied by an exponential function. Solving for the imaginary part of the energy, while retaining only first order terms in the polynomial, then leads to
\begin{align}
\Gamma_{\mathrm{asymp.}}\left(\mathcal{E}_{\mathrm{DC}}\right)=\left(1-\frac{5}{3}\mathcal{E}_{\mathrm{DC}}\right)\exp\left(-\frac{2}{3\mathcal{E}_{\mathrm{DC}}}\right)\thinspace.\label{eq:asympion}
\end{align}
The ionization rate of the monochromatic field can then again be obtained by \cref{eq:adint} using \cref{eq:asympion} for $\Gamma_{\mathrm{DC}}$. It can be seen to agree with the first two methods for weak fields.
The final comparison is made with the expression obtained by Perelomov, Popov, and Terent'ev (PPT) \cite{Perelomov1966} for the adiabatic ionization rate for the zero-range potential in a monochromatic field with amplitude $\mathcal{E}_0$
\begin{align}
\langle \Gamma _{\mathrm{PPT}}\left(\mathcal{E}_0\right) \rangle = \left(\frac{3\mathcal{E}_0}{\pi}\right)^{1/2}\exp\left(-\frac{2}{3\mathcal{E}_0}\right) \thinspace.\label{eq:Perelomov}
\end{align}
It again agrees for weaker fields but as opposed to the approximation obtained by the asymptotic analysis it overestimates the ionization rate for strong fields.
\begin{figure}[t]
	\includegraphics[width=1\columnwidth]{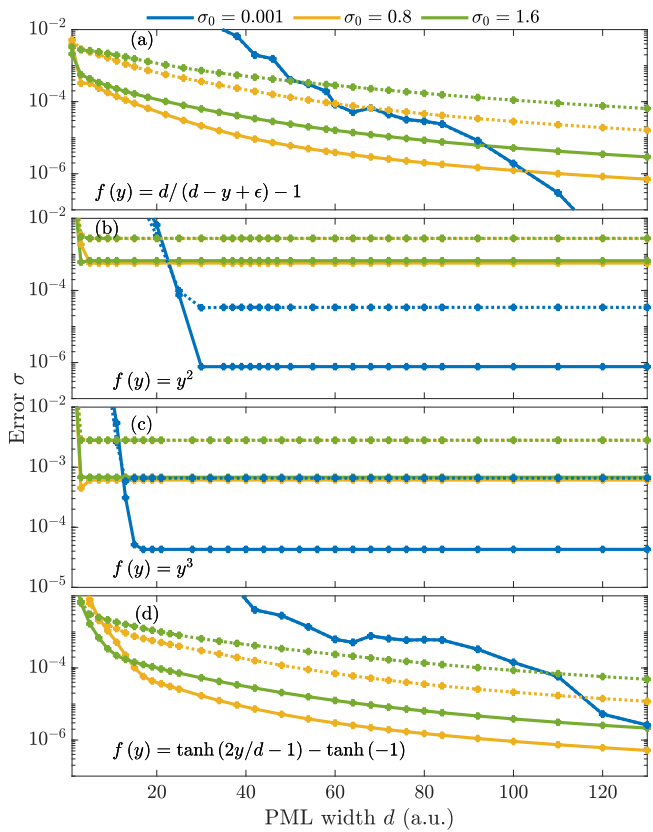}
	\caption{Error at time $t=200$ a.u. as a function of PML width for four different absorption functions and various absorption coefficients. The solid and dotted lines correspond to PML VG and LG, respectively. The parameters are $\mathcal{E}_0 = 0.1$ a.u. and $\omega=0.52$ a.u., and the field has been turned on smoothly over three optical cycles. }\label{fig:errorvsPMLHydr}
\end{figure}
\section{Long-range Potential}\label{sec:LR}
\begin{figure}[t]
	\includegraphics[width=1\columnwidth]{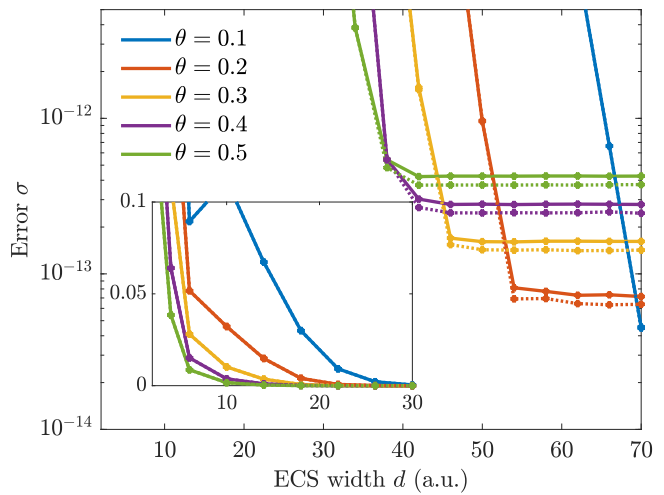}
	\caption{Error at time $t=200$ a.u. as a function of ECS width for various angles of rotation. The field parameters are the same as in \cref{fig:errorvsPMLHydr}. The inset shows the behavior for smaller widths. }\label{fig:errorvsECSHydr}
\end{figure}

\begin{figure}
	\includegraphics[width=1\columnwidth]{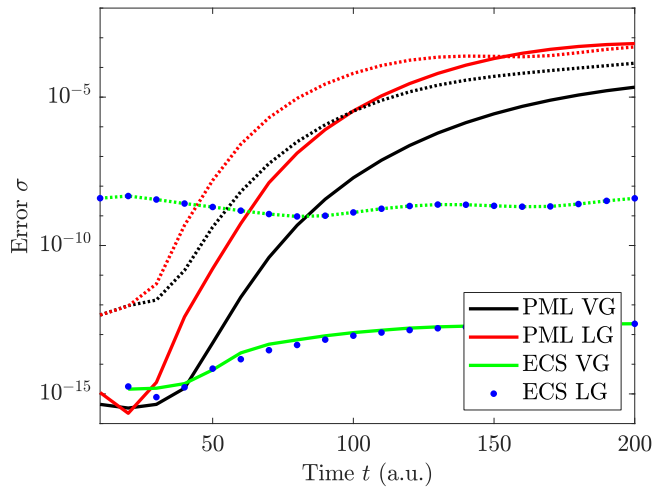}
	\caption{Error as a function of time. The absorbing boundary is located at $R_0=20$ and $R_0=10$ a.u. for the solid and dotted lines, respectively. The PML calculations have been made with $\sigma_0 = 0.001$ and the quadratic absorption function, while the ECS calculations have been made with $\theta = 0.35$. In both cases the absorption width is $d = 40$ a.u.. The field parameters are the same as in \cref{fig:errorvsPMLHydr}.   }\label{fig:errorvstimeHydr}
\end{figure}

As a final study of the behavior of PMLs in laser-matter interactions, we will look at a long-range potential, namely the one-dimensional "hydrogen atom" 
\begin{align}
	V\left(x\right) = -\frac{1}{\sqrt{x^2+2}}\thinspace.
\end{align}
This potential does not have the same cut-off spatial behavior as the short-range potential in the previous section, and thus this potential will reach into the absorbing layer. The ground-state energy of the one-dimensional hydrogen atom remarkably comes out as exactly $E_0 = -1/2\,\mathrm{a.u.}$.
First, we study the error as a function of absorption width $d$. This is shown in \cref{fig:errorvsPMLHydr} for the PML method and in \cref{fig:errorvsECSHydr} for ECS. For the PML method, we see similar trends as for the short-range potential with the exception that the errors are much larger. Whereas the PML with $\sigma_0 = 0.001$ and a square absorption function converged to an error of the order $10^{-15}$ for the short-range potential, it converges to around $10^{-6}$ for the long-range potential. It does seem, however, that one is able to obtain lower errors if the absorption width is increased and the absorption function is allowed to increase more slowly. This is indicated in \cref{fig:errorvsPMLHydr}(a), as the nearly singular function will grow slower as $d$ is increased. This leads to an interesting opportunity in implementing a non-uniform grid in the absorption region, significantly reducing the number of grid points required to describe a much wider absorption width. This was done in Ref. \cite{Weinmuller2017} for ECS and was shown to produce excellent results. \Cref{fig:errorvstimeHydr} shows the error as a function of time for the same parameters as for the short-range potential. Here it can be seen that ECS in VG and LG are again indistinguishable at later times. As is evident, the ECS scheme is preferable for a potential that reaches into the absorbing layer. 

\section{Concluding Remarks}

We have presented a simple FDTD scheme implementing PMLs to describe the dynamics of an electron in a laser field. The PML approach has been compared to the well known ECS approach, and we observe that the PML, when implemented in VG, far outperforms the ECS approach when the potential vanishes outside the scaling radius $R_0$. Conversely, the traditional ECS approach is much more efficient when implemented for a potential that reaches into the absorption domain. Upon comparing the errors introduced by ECS implemented in the two gauges, little to no difference is observed. On the other hand, when PMLs are implemented in LG, significantly larger errors can be seen. For sufficiently low frequencies, both the ECS LG and VG and the PML VG wave functions blow up, leaving PML LG as the only remaining option of the four methods. Finally, we have demonstrated that the PML implemented in LG is able to reproduce the ionization rate in the adiabatic region, where the other methods considered fail.

\begin{acknowledgments}
Useful comments from Lars Bojer Madsen are gratefully acknowledged. This work was supported by the Villum Kann Rasmussen (VKR) Center of Excellence QUSCOPE. Additionally, H.C.K. and T.G.P. are supported by the Center for Nanostructured Graphene (CNG), which is sponsored by the Danish National Research Foundation, Project No. DNRF103.
\end{acknowledgments}

\appendix
\section{Finite Difference Formulas}\label{app:FD}

The finite difference (FD) approach used in the present paper is based on the Crank-Nicolson scheme \cite{Crank1996}. It consists of a combination of the forward (explicit) and backward (implicit) Euler method and reads
\begin{multline}
	i\frac{\psi\left(x,t_{j+1}\right)-\psi\left(x,t_{j}\right)}{\Delta t}\\ = \frac{1}{2}\left[H\left(t_{j+1}\right)\psi\left(x,t_{j+1}\right)+H\left(t_{j}\right)\psi\left(x,t_{j}\right)\right]\thinspace,
\end{multline}
where $t_j = j\Delta t$ and $H\left(t_j\right)$ is the Hamilton operator at time $t_j$. What remains is to discretize the spatial derivatives. For the PML method, the kinetic term is given by
\begin{multline}
	T_{\mathrm{PML}} = -\frac{1}{2}c\left(x\right)\frac{\partial }{\partial x}c\left(x\right)\frac{\partial}{\partial x}\psi \\
	 = -\frac{1}{2}\left[c^2\left(x\right)\frac{\partial ^2\psi}{\partial x^2} + c\left(x\right)\frac{\partial c\left(x\right)}{\partial x}\frac{\partial \psi}{\partial x}\right]\thinspace.
\end{multline}
Both $c$ and its derivative are known analytically. The derivatives of the wave function are approximated using the second-order approximations
\begin{align}
\frac{\partial \psi}{\partial x}&\approx \frac{\psi_{n+1}-\psi_{n-1}}{2\Delta x} \\
\frac{\partial^2 \psi}{\partial x^2}&\approx \frac{\psi_{n+1}-2\psi_{n}+\psi_{n-1}}{\left(\Delta x\right)^2}\thinspace,
\end{align}
with $\psi_n = \psi\left(x_n\right)$ and $x_n = n\Delta x$, where $n$ is an integer running from $-N$ to $N$ for a grid with a total of $2N+1$ equidistant points.  These simple FD formulas are one of the advantages of the PML method. 

For the ECS method, on the other hand, the transformation leads to modified FD formulas. They can be derived by writing the wave function as a standard approximation using Lagrange interpolating polynomials
\begin{align}
\psi\left(\tilde{x}\right) \approx \sum_{n=-p}^{p}l_n\left(\tilde{x}\right)\psi\left(\tilde{x}_n\right)\label{eq:LagrangeWF}
\end{align}
with
\begin{align}
l_n\left(\tilde{x}\right) = \prod_{\substack{m=-p \\m\neq n}}^{p}\frac{\tilde{x}-\tilde{x}_m}{\tilde{x}_n-\tilde{x}_m}\thinspace.
\end{align}
The FD formulas at any particular point are then derived by differentiating Eq. \eqref{eq:LagrangeWF} and evaluating the result at said point, all the while keeping in mind Eq. \eqref{eq:ECStrans}. That is, 
\begin{align}
\tilde{x}_n =\begin{cases}
	x_n \quad &\text{for } x_n\leq R_0\\
	e^{i\theta}\left(x_n\pm R_0\right)\mp R_0 \quad &\text{for } \mp x>R_0\thinspace,
\end{cases}	 
\end{align}
where $x_n$ is the equidistant grid described above. Here we use $p=1$, which leads to the standard FD formulas for $\left|x\right|<R_0$. Outside the scaling radius we simply pick up a complex phase factor
\begin{align}
\frac{d\psi}{dx}&\approx \frac{e^{-i\theta}}{2\Delta x}\left(\psi_{n+1}-\psi_{n-1}\right)\\
\frac{d^2\psi}{dx^2}&\approx \frac{e^{-i2\theta}}{\left(\Delta x\right)^2}\left(\psi_{n-1}-2\psi_{n}+\psi_{n+1}\right)\thinspace,
\end{align}
while at the scaling radius we have the non-symmetric formulas
\begin{multline}
\frac{d\psi}{dx} \approx \frac{1}{\Delta x}\left[-\frac{e^{\pm i\theta}}{e^{i\theta}+1}\psi_{n-1}\pm\left(1-e^{-i\theta}\right)\psi_n\right.\\\left.+\frac{e^{\mp i\theta}}{e^{i\theta} +1}\psi_{n+1}\right]\quad \mathrm{for }\, x= \pm R_0
\end{multline}
\begin{multline}
\frac{d^2\psi}{dx^2}\approx \frac{1}{\left(\Delta x\right)^2}\left[\frac{2}{e^{i\theta}+1}\psi_{n\mp1}-2e^{-i\theta}\psi_n \right.\\\left. + \frac{2e^{-i\theta}}{e^{i\theta}+1}\psi_{n\pm1}\right]\quad \mathrm{for }\, x= \pm R_0
\end{multline}
As was discussed in Ref. \cite{McCurdy2004}, the ECS FD formulas are $O\left[\left(\Delta x\right)^2\right]$ for $x\neq \pm R_0$ but only $O\left(\Delta x\right)$ for $x= \pm R_0$. For all calculations in the present paper, we have used $\Delta x = 10^{-2}\,\mathrm{a.u.}$ and $\Delta t = 10^{-3}\,\mathrm{a.u.}$.

\bibliography{litt}

\end{document}